\documentclass[onecolumn]{aastex6}

\newcolumntype{?}[1]{!{\vrule width #1}}

\collaborationName{Friends of AASTeX}
\usepackage{verbatim}
\usepackage{amsmath}
\usepackage{rotating}
\DeclareMathOperator*{\mini}{minimize}

\begin{document}

\title{Towards the ICRF3: astrometric comparison of the USNO 2016A VLBI solution with ICRF2 and Gaia DR1}

\author{Julien Frouard\altaffilmark{1}, Megan C. Johnson\altaffilmark{2}, Alan Fey, Valeri V. Makarov and Bryan N. Dorland}
\affil{United States Naval Observatory (USNO)\\
3450 Massachusets Ave NW, Washington, DC 20392, USA}

\altaffiltext{1}{julien.frouard.ctr@navy.mil}
\altaffiltext{2}{megan.johnson@navy.mil}

\begin{abstract}

The VLBI USNO 2016A (U16A) solution is part of a work-in-progress effort by USNO towards the preparation of the ICRF3. Most of the astrometric improvement with respect to the ICRF2 is due to the re-observation of the VCS sources. Our objective in this paper is to assess U16A's astrometry. A comparison with ICRF2 shows statistically significant offsets of size 0.1 mas between the two solutions. While Gaia DR1 positions are not precise enough to resolve these offsets, they are found to be significantly closer to U16A than ICRF2. In particular, the trend for typically larger errors for Southern sources in VLBI solutions are decreased in U16A. Overall, the VLBI-Gaia offsets are reduced by 21\%. The U16A list includes 718 sources not previously included in ICRF2. Twenty of those new sources have statistically significant radio-optical offsets. In two-thirds of the cases, these offsets can be explained from PanSTARRS images. 

\end{abstract}

%\keywords{editorials, notices --- miscellaneous --- catalogs --- surveys}



\section{Introduction}

The ICRF2 catalog \citep{fey+2015} is the current realization of the International Celestial Reference Frame (ICRF) since 2010. This paper discusses some of the work done in preparation of the future iteration of the ICRF, the ICRF3. Most of the current improvement discussed in this paper comes from the re-observation of the VCS sources (VCS-II, see \citet{Gordon+2016}). 

In the optical domain, the Gaia mission already produced astrometrically unmatched positions of the ICRF2 optical counterparts in its first data release \citep{lindegren+2016}. In particular, \cite{mignard+2016} investigated the distribution of offsets between ICRF2 and the Gaia Auxiliary Quasar Solution; they found that the bulk of the sources have a scatter of  $\sim 1.5-1.8$ mas in the coordinates for the whole sample and   $\sim 0.7$ mas for the defining sources. They also found a significant bias of -0.1 mas in declination (in the sense Gaia - ICRF2) for the defining subset.  

While the ICRF2 is built uniquely on S/X observations, recent efforts with observations in the K-band give solutions closer to GAIA DR1 \citep{jacobs2017}, and are thus already a significant improvement to the ICRF2, even with the limited number of sources observed up to now. This makes the contribution of K-band observations highly compelling for the future of the ICRF.

Many works have discussed the presence of offsets between the radio and optical positions (see for example \cite{assafin2013,orosz2013,zacharias2014}). The offsets can give some information on the intrinsic structure of the source, such as the presence of a jet structure \citep{kovalev2017}. Several papers recently investigated the statistically significant radio-optical offsets using radio positions and the larger secondary solution of the Gaia DR1 catalog \citep{petrov2017a,makarov2017}. \cite{makarov2017} investigated the statistically significant offsets between ICRF2 and Gaia DR1 and used PanSTARRS \citep{chambers2016} images to find, when possible, the underlying reason; close doubles (actual binaries or chance alignements with field stars), source confusion or extended objects. Knowledge of the large offsets is important as they can complicate the astrometric adjustment of large catalogs to the ICRF (see \cite{berghea2016} for a discussion of this issue).

This paper discusses the astrometric differences between the work-in-progress toward ICRF3 developed at USNO, called USNO 2016A (hereafter named U16A), the ICRF2, and the Gaia DR1 catalog.

\section{The USNO 2016A Global CRF Solution}

The United States Naval Observatory is one of the VLBI Analysis Centers for the International VLBI Service (IVS) and as such, is responsible for producing Celestial and Terrestrial Reference Frame (CRF and TRF, respectively) products for the community.  Here, we present the U16A global solution for the CRF derived using observations that date back to the onset of the astrometric/geodetic VLBI in 1979.  Data were acquired at standard wavelengths of 13 and 3.6 cm simultaneously; this allows for accurate calibration of the wavelength-dependent propagation delays introduced by the Earth's ionosphere. VLBI astrometry uses multiple channels within each band and applies a least-squares analysis method in order to produce precise group delays. Typically, VLBI observing sessions occur over a 24-hour period, which is necessary to separate the parameters for polar motion and nutation.  The U16A solution was determined using the CALC/SOLVE software currently distributed by Goddard Space Flight Center (GSFC).  For a comprehensive description of the least-squares method used for deriving source positions from group delay observations, we refer to \cite{ma86}.

The U16A solution contains a total of 4129 sources, 295 of which are part of the ``defining sources'' category of VLBI sources.  These 295 objects have been observed with the highest number of observing sessions and thus, have the most constrained positions in the solution.  There are 2195 sources that are part of the VLBA Calibrator Survey (VCS) as presented in \cite{beasley02}, which have since been re-observed. These VCS sources are among the least observed objects in the solution with a median number of observing sessions of only two and therefore, have the least constrained positions.  The remaining objects are a mixture of new and other, non-VCS sources.  The U16A solution contains 3411 objects from the ICRF2.  Three ICRF2 sources are absent from the U16A solution: SN1993J, VELA, and LANA. Table \ref{tab5} summarizes the different sources and the number of times the objects have been observed.  In the next section, we compare the differences in position between the U16A solution and the ICRF2.

\begin{table}[h]
\begin{center}
\begin{tabular}{l?{0.25mm}r?{0.25mm}r|r|r?{0.25mm}r|r|r}
Type of source		& Number  	& \multicolumn{3}{c}{Number of sessions} 	& \multicolumn{3}{c}{Number of observations (delays)}	\\
			&   		& min&median&max  	& min&median&max	\\
\hline
Defining 		& 295		& 27&190&4398		& 104&4376&385497\\
VCS 			& 2195		& 1&2&34		& 12&113&674 \\
Non-VCS 		& 921		& 1&16&3830		& 3&214&234292\\
New 			& 718		& 1&2&18		& 3&59&864\\
\textbf{Total} 		& \textbf{4129}	& \textbf{1}&\textbf{2}&\textbf{4398}	& \textbf{3}&\textbf{111}&\textbf{385497}
\end{tabular}
\end{center}
\caption{Statistics of the U16A solution}
\label{tab5}
\end{table} 

\section{Comparison of the U16A solution with ICRF2}

\begin{figure}[h]
\centering
\includegraphics[width=5in]{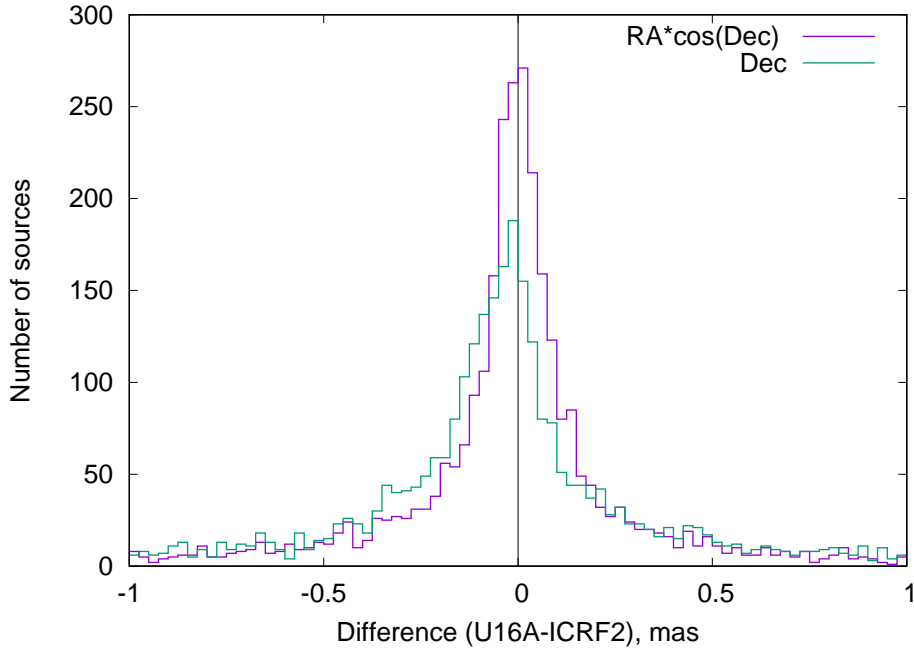}
\caption{Position differences in mas, in the sense U16A-ICRF2, restricted to the range [-1,1] mas.}
\label{fig2}
\end{figure}

Figure \ref{fig2} shows the differences U16A-ICRF2 for both coordinates. We can already notice a clear systematic difference in the declination component. Table \ref{tab1} gives the median differences in components and absolute offsets. The numbers between parentheses are the confidence intervals of the median, computed with bootstrap resampling \citep{efrontibshirani1994}, using the BC$_a$ (Bias-Corrected and accelerated) intervals \citep{diciccioefron1996} with 95\% coverage and 20,000 bootstrap samples. The advantage of using the bootstrap here is to i) provide a statistical uncertainty for the median estimator, and ii) compute that uncertainty without assuming any error distribution. We will study the consistency of the U16A formal errors in a later section.

\begin{table}[h]
\begin{center}
\begin{tabular}{l|rc|rc|rc|rc}
			& \multicolumn{2}{c}{295 defining}  	& \multicolumn{2}{c}{2191 VCS} 		& \multicolumn{2}{c}{921 Non-VCS} 	& \multicolumn{2}{c}{\textbf{3407 sources}}\\
\hline
RA*cos(Dec) 		& -9 	&(-24, 12)			&  -31& (-49, -7)			& -3 &(-19, 4)				& \textbf{-17} &\textbf{(-25, -4)}\\
Dec 			& -45	& (-60, -35)			& -27 &( -55, -6)			& -42& (-52, -35) 			& \textbf{-39} &\textbf{(-50, -31)}\\
Angular separation	& 90 	&(79, 100) 			& 572 &(517, 618) 			& 141 &(129, 150)  			& \textbf{306} &\textbf{(286, 328)}
\end{tabular}
\end{center}
\caption{Median differences in position in the sense U16A-ICRF2, in $\mu$as. The number between parentheses are bootstrap BC$_a$ confidence intervals with 95\% coverage. The median differences in both coordinates represent the systematic offset between the solutions, while the median angular separation gives a sense of the scatter of the absolute displacement.}
\label{tab1}
\end{table} 

The median of the absolute offsets between the two solutions is $\sim$ 0.3 mas. However the differences are heterogeneous when we consider the different subsets and components. There is a clear difference in declination for the defining and non-VCS sources between the two catalogs, while the VCS subset seems less affected. As noted above, a 0.1 mas offset was similarly reported by \cite{mignard+2016} in a comparison between Gaia AQS and ICRF2. In comparison, the differences in RA*cos(Dec) are much less sharp, and the only statistically significant difference concerns the VCS sources.

\begin{figure}[h]
\centering
\includegraphics[width=3.4in]{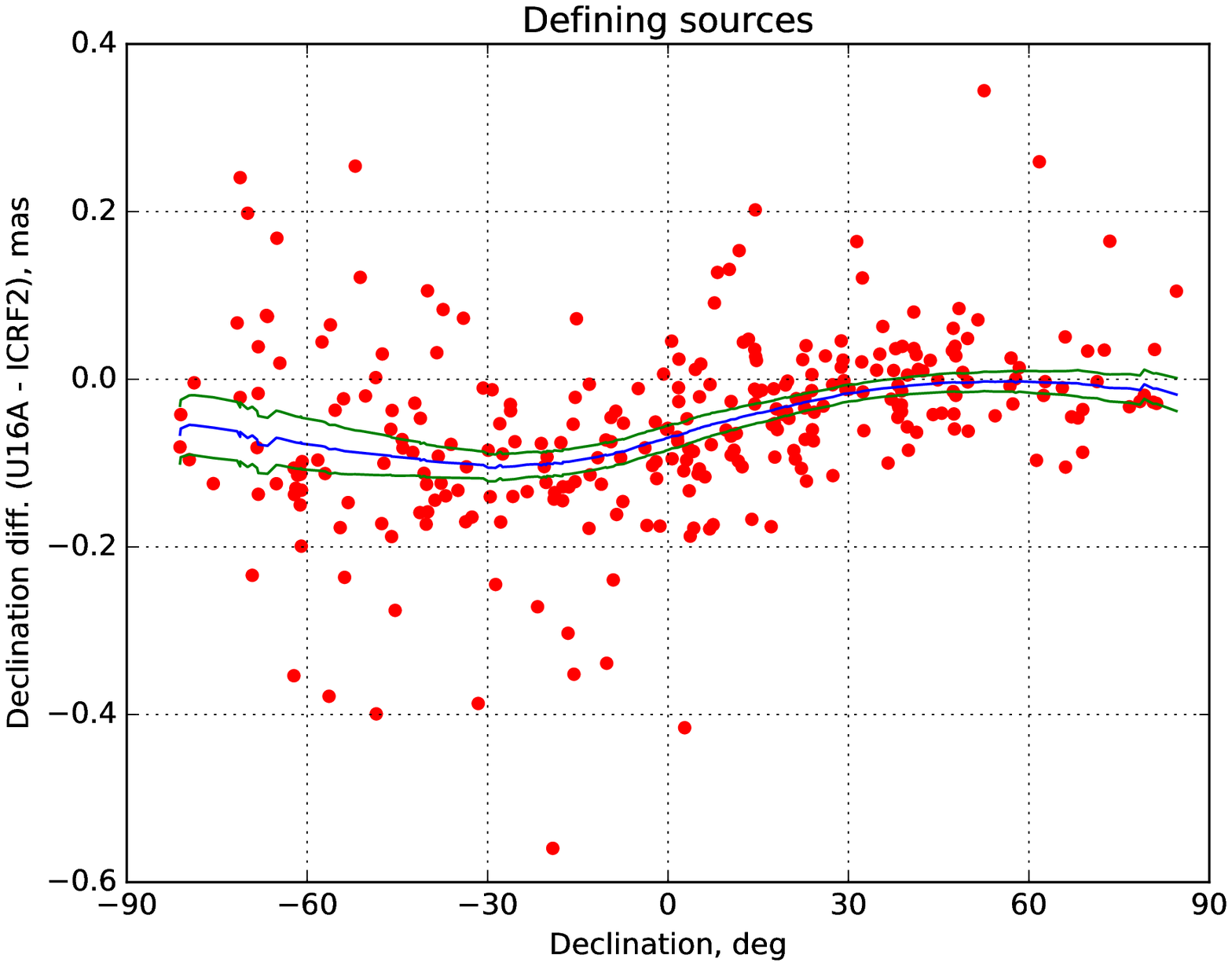}
\includegraphics[width=3.4in]{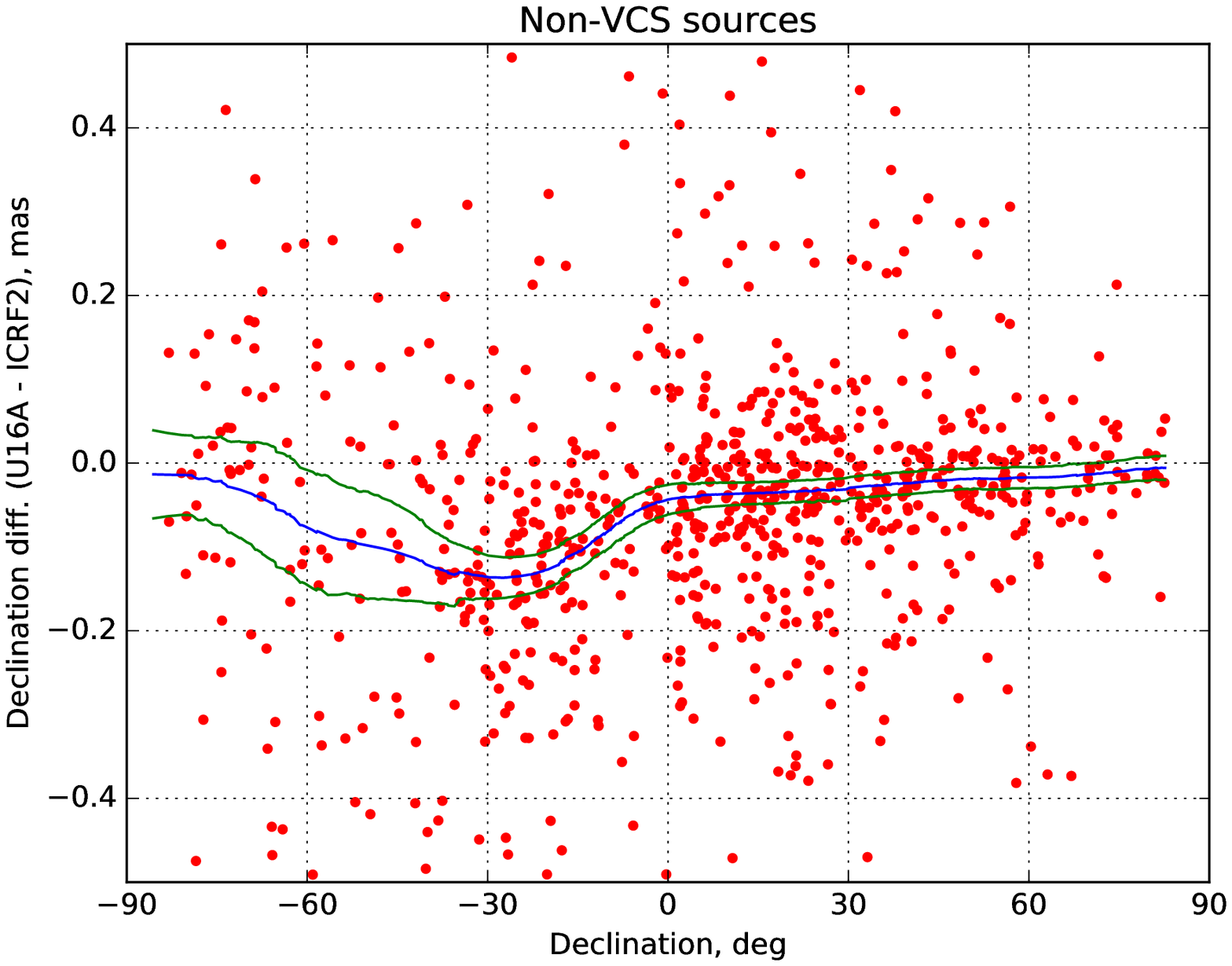}
\includegraphics[width=3.4in]{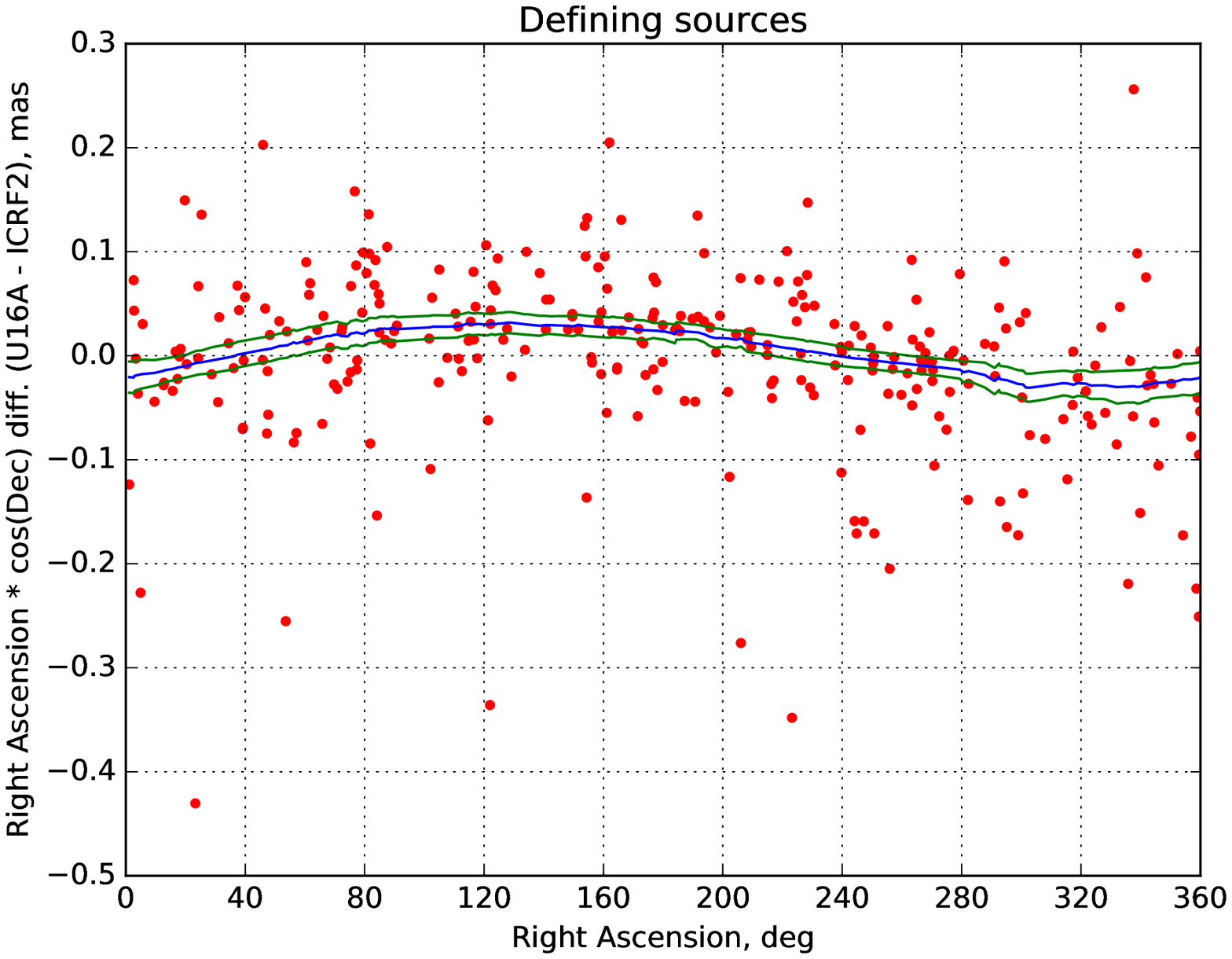}
\caption{Position differences in mas, in the sense U16A-ICRF2. The blue curve is a robust local estimate of the mean, while the green curves indicate its 95\% confidence interval. See appendix A for details.}
\label{fig3}
\end{figure}

The behavior of the offsets with respect to the right ascension and declination is shown on figure \ref{fig3}, where we only show the trends that indicate a significant, non-linear behavior with respect to the components. The blue line indicates a local robust average, while the green lines represent a 95\% confidence interval computed with bootstrap resampling. Details on the statistical method are given in the appendix. One of the main features is a clear negative systematic offset in the declination, mostly affecting the southern declinations for the defining and non-VCS subsets and peaking at $\delta \sim$ - 30 deg.

Another distinct, though smaller, feature is a sinusoidal pattern of the right ascension offset with respect to that same coordinate for the defining sources.
 
Finally, we note that there are four outliers with absolute position offsets U16A-ICRF2 larger than 150 mas in at least one coordinate\footnote{These sources are not present in the Gaia AQS catalog.}. These are IVS 0114-211 (offset = 4484 mas), IVS 1936+095 (offset = 1084 mas), IVS 0802-170 (offset = 379 mas), and IVS 1820-274 (offset = 303 mas). These large offsets are likely due to the improved astrometric quality\footnote{Only IVS 0802-170 is present in the Gaia DR1 main catalog. It gives a good example of U16A's improved astrometry; its Gaia position being very close to U16A (0.57 mas) compared to ICRF2 (379.24 mas).}, resulting from an increase in the number of delays from the ICRF2. The ICRF2 contained only 3 delay measurements per source, whereas the present U16A solution includes between 12 and 54 delay measurements per source.

\subsection{Global rotation and glide between the different frames}

It is also of interest to examine whether statistically significant large-scale patterns are present between the three frames. We applied a vector spherical harmonic (VSPH) expansion of the offsets in RA and Dec following the methods outlined in \cite{makarov2007} and \cite{mignard2012}. The VSPH expansion decomposes the offsets into a set of spheroidal and toroidal harmonics; the first degree of which corresponds to a global rotation and ``glide'', where the last term, coined by \cite{mignard2012}, represents a pattern in which the offsets are all directed to a specific point on the celestial sphere. The VSPH expansion, being formally infinite, is usually computed up to a given degree $l_{max}$. Which harmonics are statistically significant or useful to describe the data must then be decided. Here we computed the expansion for different maximum degrees ($l_{max} = 1...7$), and selected the $l_{max}$ which minimizes the normalized Sum of the Squared Residuals $SSR / (2N - 2p)$, where $N$ is the number of sources and $p$ is the number of spherical harmonics considered. Choosing $l_{max}$ on the basis of the BIC criterion \citep[Bayesian Information Criterion, see][]{schwarz1978} gave identical results. Different procedures like cross-validation or bootstrap could also be used for that task. As a sidenote, we take notice that since the VSPH expansion is a useful tool to analyze sky-correlated patterns in astrometry, a deeper consideration of what is the ``adequate'' $l_{max}$, weighting the different techniques of model selection, would be interesting. 

The computations were made with a weighted least squares solving equation (30) in \cite{mignard2012} up to the maximum degree $l_{max}$, and taking into account the covariance errors. The outliers were removed by discarding the offsets located at more than 3 $\times$ MAD\footnote{MAD is the (normalized) Median Absolute Deviation from the median : MAD(a) = median($|a_i$ - median(a)$|$) $\times$ 1.4826. The numerical factor renders the MAD equal to the standard deviation at the standard normal distribution.} from the sample median. The basic rotation and glide between U16A and ICRF2 are reported in table \ref{tab10}. These are the simplest global offset patterns and correspond to the lowest degree harmonics (specifically, $l=1$ with $m=0,\pm 1$). The components of the rotation and glide vectors are obtained from the spherical harmonics using equations (62-63) and (67-68) from \cite{mignard2012}. We do not report the offset patterns corresponding to higher-degree coefficients. The number of sources $N_{sources}$ considered is indicated in the table. The component values shown in the table correspond to a computation using the $l_{max}$, also indicated in the table. 

\begin{table}
 \begin{center}
\begin{tabular}{llllcccccc}
		&  & $l_{max}$ 	& $N_{sources}$ & \multicolumn{3}{c}{Rotation} & \multicolumn{3}{c}{Glide}\\
		&  &   		&  	 	& x 			& y 			& z 			& x 			& y 			& z\\
\hline
\hline
U16A - ICRF2 	&  & 2 		& 2433 		& -3 $\pm$ 4		& -10 $\pm$ 4		&  0.7 $\pm$ 3		& $\mathbf{-19 \pm 4}$	& $\mathbf{-29 \pm 4}$	& $\mathbf{-71 \pm 4}$\\
U16A - Gaia AQS	&  & 1 		& 1794 		& -41 $\pm$ 14 		& $\mathbf{-54 \pm 12}$ & -4 $\pm$  13  	&-23  $\pm$ 13  	& $\mathbf{60  \pm 12}$  & $\mathbf{78 \pm 11}$\\
ICRF2 - Gaia AQS&  & 1 		& 1728 		& $\mathbf{-54 \pm 16}$ & $\mathbf{-55 \pm  14}$&  16$\pm$ 15		& -13 $\pm$ 15		&$\mathbf{ 88 \pm 15 }$	& $\mathbf{155 \pm 14 }$
 \end{tabular}
 \end{center}
\caption{Components of the global rotation and glide between the catalogs, in $\mu$as. The bold font denotes the values with a 3-sigma significance.}
\label{tab10}
\end{table}

\begin{figure}[h]
\centering
\includegraphics[width=3.2in]{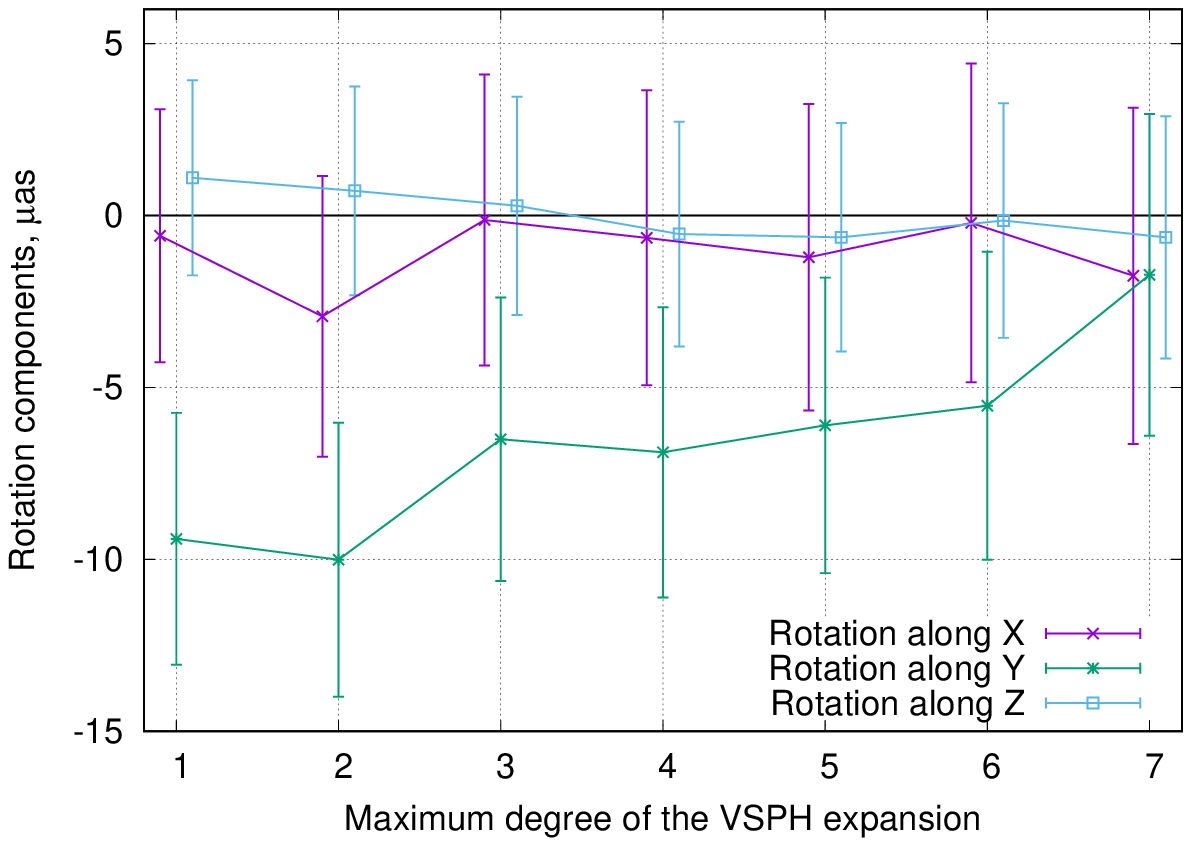}
\includegraphics[width=3.2in]{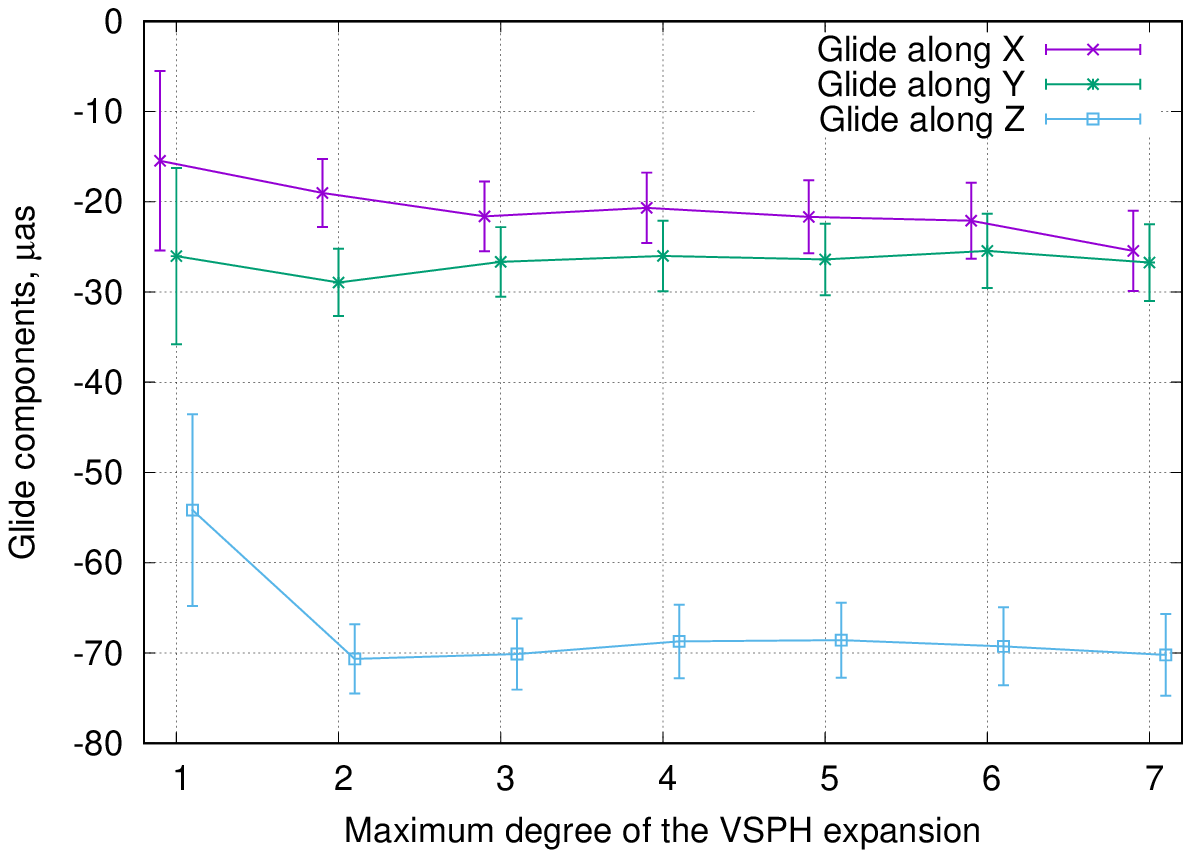}
\caption{Rotation and glide components resulting from the VSPH expansion of the U16A - ICRF2 offsets. The values are slightly shifted along the horizontal for better readability.}
\label{fig11}
\end{figure}

The rotation and glide components of the offsets between the U16A and ICRF2 positions are shown in Figure \ref{fig11}, for different values of $l_{max}$. They do not show a statistically significant rotation (at the 3-sigma level). The main feature is the highly significant glide present in each coordinate, and its large negative $z$-component is consistent with the negative offsets in declination seen in table \ref{tab1}. These two results also hold for computations using $l_{max} = 2...7$, and are thus robust in that regard.

Where do the U16A-ICRF2 systematic differences seen in figure \ref{fig3} come from? A definitive answer is unclear at this point, but as shown by \cite{ma16}, similar systematic differences have been linked to observations from stations belonging to the AUST network, which started observing in 2010. The authors showed that removing the observations from the AUST stations - as well as those from the KATH12M or HOBART12 stations - reduced some of those systematic differences. A systematic error in ICRF2 still cannot be ruled out.

\section{Comparison of the VLBI positions with the Gaia Auxiliary Quasar Solution}

Can a comparison with the Gaia positions help in identifying which solution is causing the offsets shown in the last section? In this section we compare the U16A and ICRF2 lists to the Gaia auxiliary quasar solution \citep{mignard+2016} (hereafter called Gaia AQS). In opposition to the Gaia DR1 main catalog, the positions of the quasars in Gaia AQS were computed with the assumption that their proper motions were negligible. The median of the position differences between the two Gaia catalogs is 0.76 mas.  

\begin{figure}[h]
\centering
\includegraphics[width=5in]{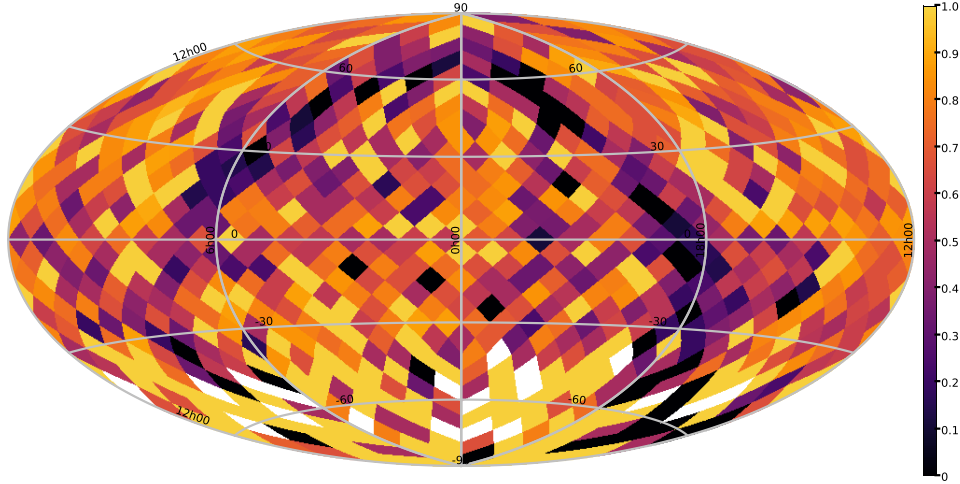}
\caption{Equatorial map (Aitoff projection) showing the ratio (number of U16A sources with a Gaia DR1 counterpart) / (number of U16A sources) in pixels of $\sim$ 54 deg$^2$. Low (darker) values indicate a scarcity of Gaia counterparts. White pixels correspond to regions without data at this resolution.}
\label{fig7}
\end{figure}

Before investigating the offsets between the catalogs, we can look at the distribution of the Gaia counterparts in RA and Dec. We refer to \cite{mignard+2016} for the problem of the selection of the optical counterparts of the ICRF2 objects in Gaia. Gaia AQS contains 2191 ICRF2 objects (64\% of the ICRF2). We matched U16A with the Gaia DR1 main catalog with the same criteria as \cite{mignard+2016}, and partitioned the sky with Healpix \citep{gorsky+2005} into pixels of size 53.7 deg$^2$ (corresponding to Healpix's \textit{Nside} = 8). We then computed the ratio (number of U16A sources with a Gaia counterpart) / (number of U16A sources) in each pixel. The result is shown in figure \ref{fig7}. Low values indicate the location where the Gaia counterparts are missing; these are mostly along the galactic plane. 

Table \ref{tab7} shows the median differences U16A-Gaia AQS and ICRF2-Gaia AQS. The latter comparison was investigated in detail by \cite{mignard+2016}. We repeat some of their computations for the sake of exploring the improvement intrinsic to U16A. The formal errors of U16A and ICRF2 are displayed in table \ref{tab9}. For comparison, the median formal errors of Gaia AQS are 626 $\mu$as (in RA) and 564 $\mu$as (in Dec).

\begin{table}[h]
\begin{center}
\begin{tabular}{ll|rc|rc|rc|rc}
	      &				& \multicolumn{2}{c}{262 defining}  & \multicolumn{2}{c}{1289 VCS} 	& \multicolumn{2}{c}{640 Non-VCS} 		& \multicolumn{2}{c}{\textbf{2191 sources}}\\
\hline
	      &RA*cos(Dec) 		& -104 &(-277, 12)	&  -108	&(-169, -27)	& -171 &(-254, -83)	& \textbf{-121} &	 \textbf{(-173, -76)}\\
U16A-Gaia AQS &Dec 			& 77 &(11, 148)		& -32 &(-90, 30)	&  36 &(-11, 91)	& \textbf{13} & \textbf{(-31, 48)}\\
	      &Angular separation 	&  584 &(489, 652)	&  1009 &(938, 1073)	&  897& (796, 991) 	& \textbf{916} & \textbf{(865, 961)}\\
\hline
\hline
	      &RA*cos(Dec) 		& -84& (-199, 10)	&  -4 &(-109, 98)	& -113 &(-221, -39)	& \textbf{-62} &	\textbf{(-124, 3)}\\
ICRF2-Gaia AQS&Dec 			& 135 &(88, 236)		& 25& (-74, 86)		&  108& (30, 176)	& \textbf{69} &\textbf{	(29, 111)}\\
	      &Angular separation 	&  601 &(491, 666)	&  1346 &(1325, 1551)	&   950	&(818, 1034)	& \textbf{1159} &	\textbf{(1097,  1225)}
\end{tabular}
\end{center}
\caption{Median differences in position in the sense U16A-Gaia AQS and ICRF2-Gaia AQS, in $\mu$as, with bootstrap BC$_a$ confidence intervals with 95\% coverage.}
\label{tab7}
\end{table} 

Overall, U16A is found to be significantly closer to Gaia AQS than ICRF2. This is mostly apparent for the VCS subset, where the offset has been reduced by $\sim$ 25 \%, compared to just a few percents for the two other subsets. This demonstrates the improvement obtained by the re-observation of the VCS sources. As shown by the size covered by the confidence intervals, the overall scatter is also much reduced; the offset of the 2191 sources has MAD(U16A-Gaia) = 768 (724, 820) $\mu$as, while MAD(ICRF2-Gaia) = 1099 (1003, 1190) $\mu$as. This is despite a larger scatter in RA*cos(Dec) for the defining sources in U16A.

Table \ref{tab4} shows the median differences between U16A and Gaia AQS limited to sources with $\geq$ 10 sessions. The positions of these sources are to be considered more precise. All the defining sources met this requirement, while only a few VCS sources did (we did not give any statistics for those). The larger scatter (i.e. the size of the bootstrap confidence interval) observed in RA*cos(Dec) is due to the defining sources, which now account for 262/657 $\sim$ 40\% of this sample.

\begin{table}[h]
\begin{center}
\begin{tabular}{l|rc|l|rc|rc}
		&  \multicolumn{2}{c}{262 defining}   	& 5 VCS 	&  \multicolumn{2}{c}{390 Non-VCS} 	& \multicolumn{2}{c}{\textbf{657 sources}}\\
\hline
RA * cos(Dec) 	&  -104 &(-277, 12)			&  N.A.		& -170& (-242, -63)			& \textbf{ -141	}&\textbf{(-219, -68)}\\
Dec 		& 77 	&(11, 148)			&  N.A.		&  39 &(-11, 98)			& \textbf{52} &\textbf{(9, 92)}\\
Absolute offset	&  584 	&(489, 652)			&  N.A.		& 703 &(606, 814)			& \textbf{644} &\textbf{(589, 702)}
\end{tabular}
\end{center}
\caption{Median differences in position in the sense U16A-Gaia AQS, in $\mu$as, with bootstrap confidence intervals with 95\% coverage. This concerns sources with $\geq$ 10 sessions. The statistics for the defining subset are identical to those in table \ref{tab7}, while the small number of VCS sources does not allow meaningful statistics.}
\label{tab4}
\end{table} 

\subsection{Scatter depending on the declination}

We investigated the systematic differences of U16A and ICRF2 with the declination, by computing the median slope of the angular separation with the Thiel-Sen method \citep{theil1950,sen1968}. We should expect to see a negative slope (i.e. larger offsets for Southern sources) due to the greater number of observing stations in the Northern Hemisphere, and the fact that Southern sources are typically observed from the Northern Hemisphere. This last feature results in small spatial sky coverage, which produces elongated uv-plane projections and highly elliptical beam patterns.

\begin{figure}[h]
\centering
\includegraphics[width=3.4in]{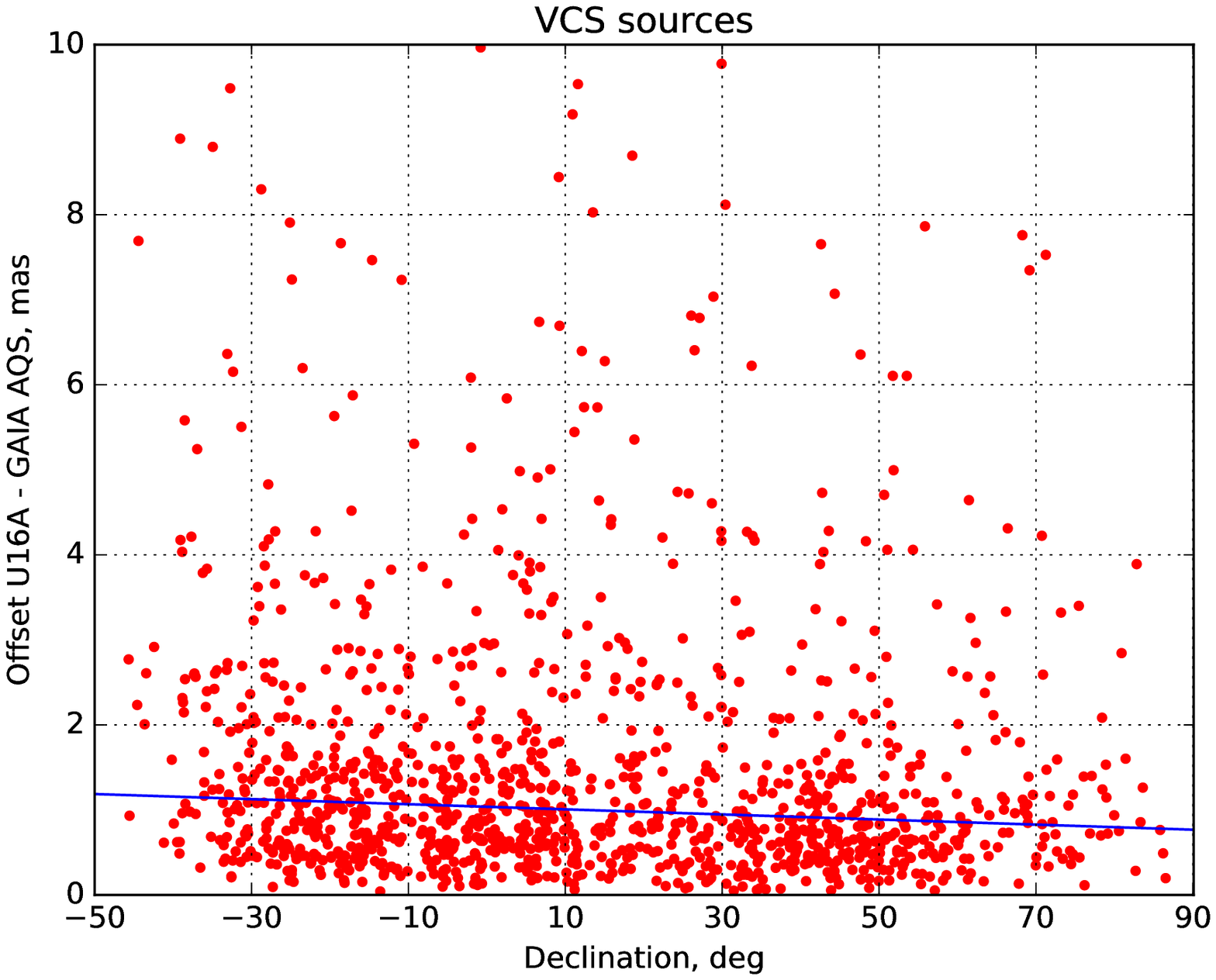}
\includegraphics[width=3.4in]{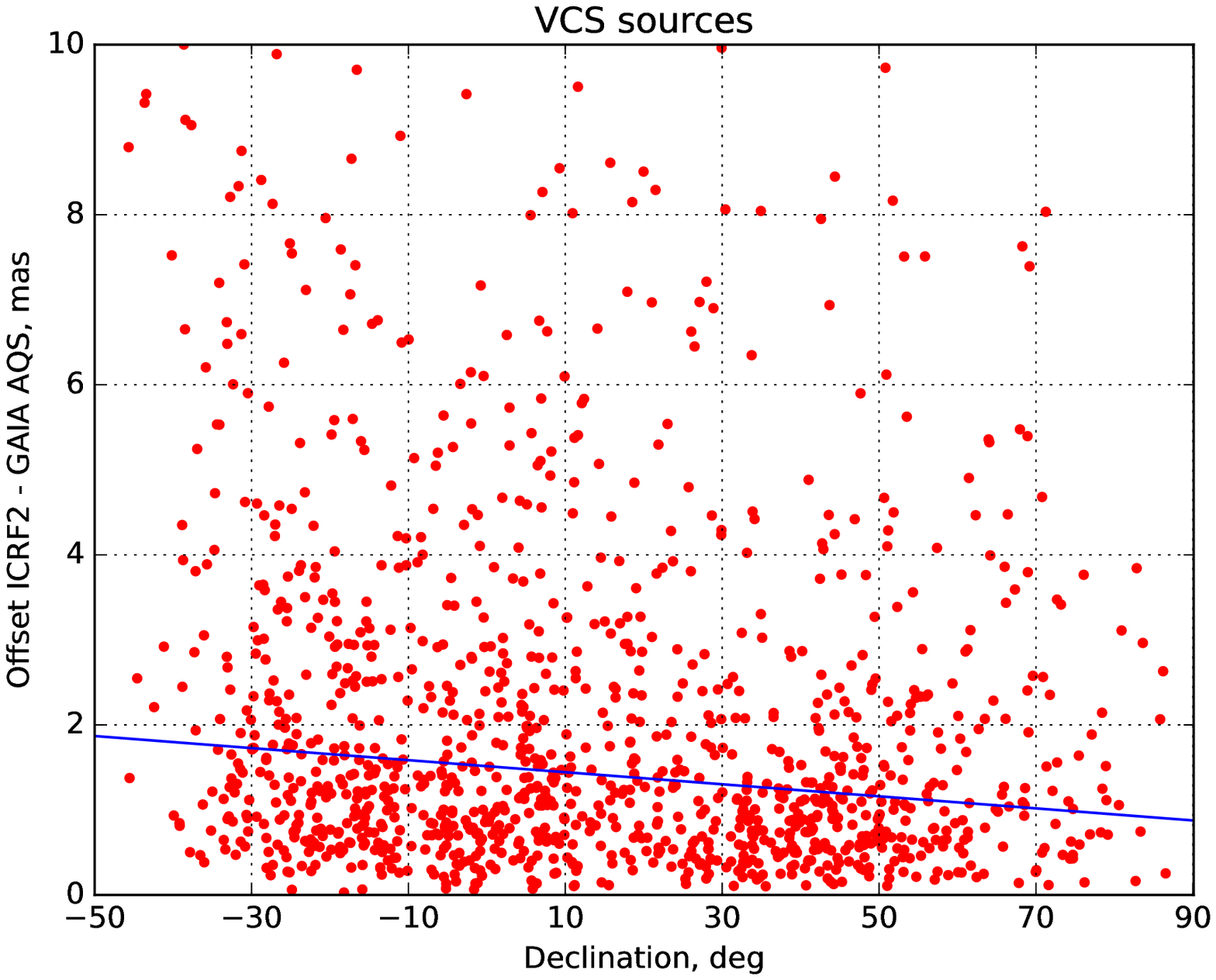}
\caption{Absolute offsets of U16A and ICRF2 relative to Gaia AQS for the VCS subset, in mas. The Thiel-Sen line (see table \ref{tab6}) is also shown.}
\label{fig12}
\end{figure}

The results are listed in table \ref{tab6} and shown in figure \ref{fig12}. The subsets have different results. The most important improvement brought by U16A concerns the VCS subset, which has the largest systematic difference in declination. The negative slope is significantly reduced by U16A, as well as its scatter. The defining subset has no statistically significant slope, both in U16A and ICRF2. The non-VCS subset does have a significant negative slope, very similar with U16A and ICRF2.

\begin{table}[h]
\begin{center}
\begin{tabular}{l|l|l|l|l}
			& 262 Defining  	& 1289 VCS 		& 640 Non-VCS 		& \textbf{2191 sources}\\
\hline
Offset U16A - Gaia AQS	&  0.51 (-0.83, 1.90)	&  -3.00 (-4.44, -1.59)	& -1.75 (-3.39, -0.29)  	& \textbf{-1.49 (-2.40, -0.59)}\\
Offset ICRF2 - Gaia AQS	&  0.29 (-1.09, 1.75)	&  -7.09 (-9.54, -4.75)	& -1.87 (-3.51, -0.35)  	& \textbf{-2.39 (-3.63, -1.19)}
\end{tabular}
\end{center}
\caption{The slope of the absolute offsets with declination, in $\mu$as.deg$^{-1}$. The slope and the 95\% coverage confidence intervals (indicated between parentheses) are determined with the Thiel-Sen method.}
\label{tab6}
\end{table} 

\subsection{Consistency of the U16A formal errors}

\begin{figure}[h]
\centering
\includegraphics[width=3.4in]{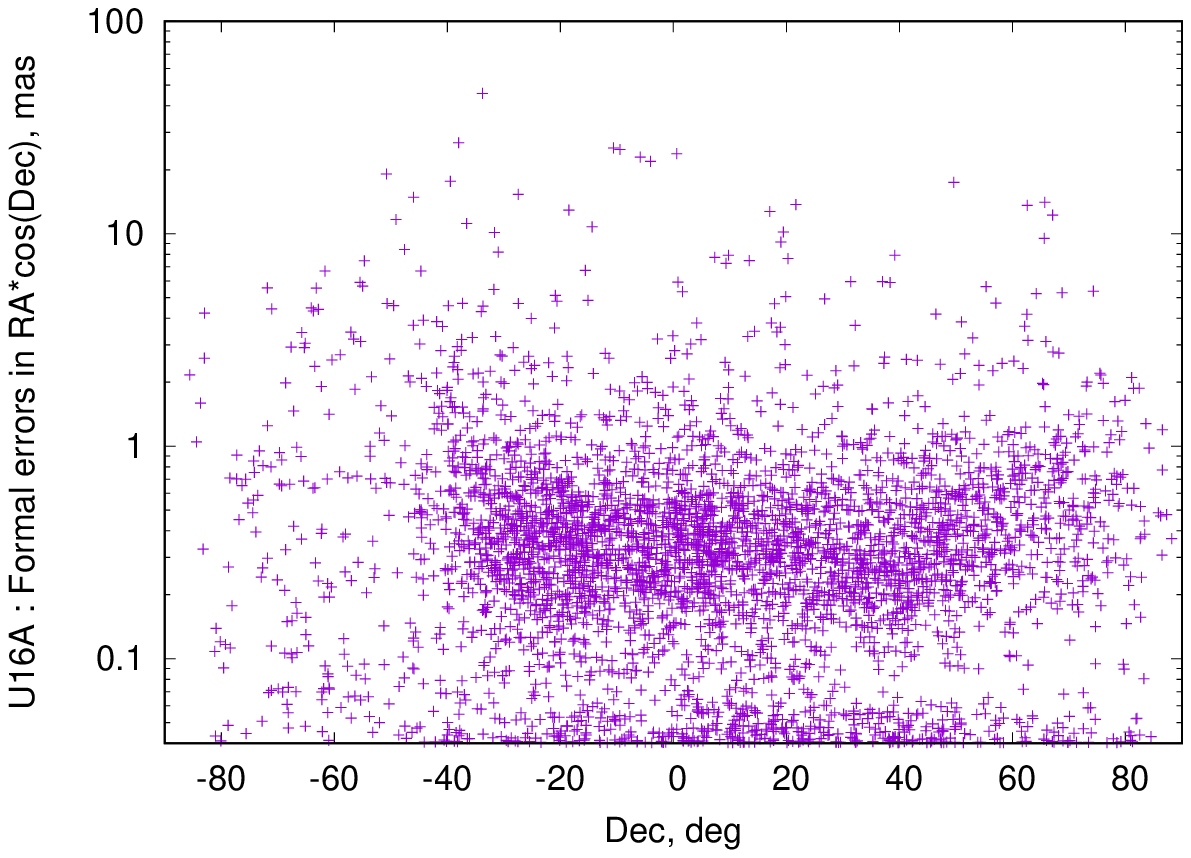}
\includegraphics[width=3.4in]{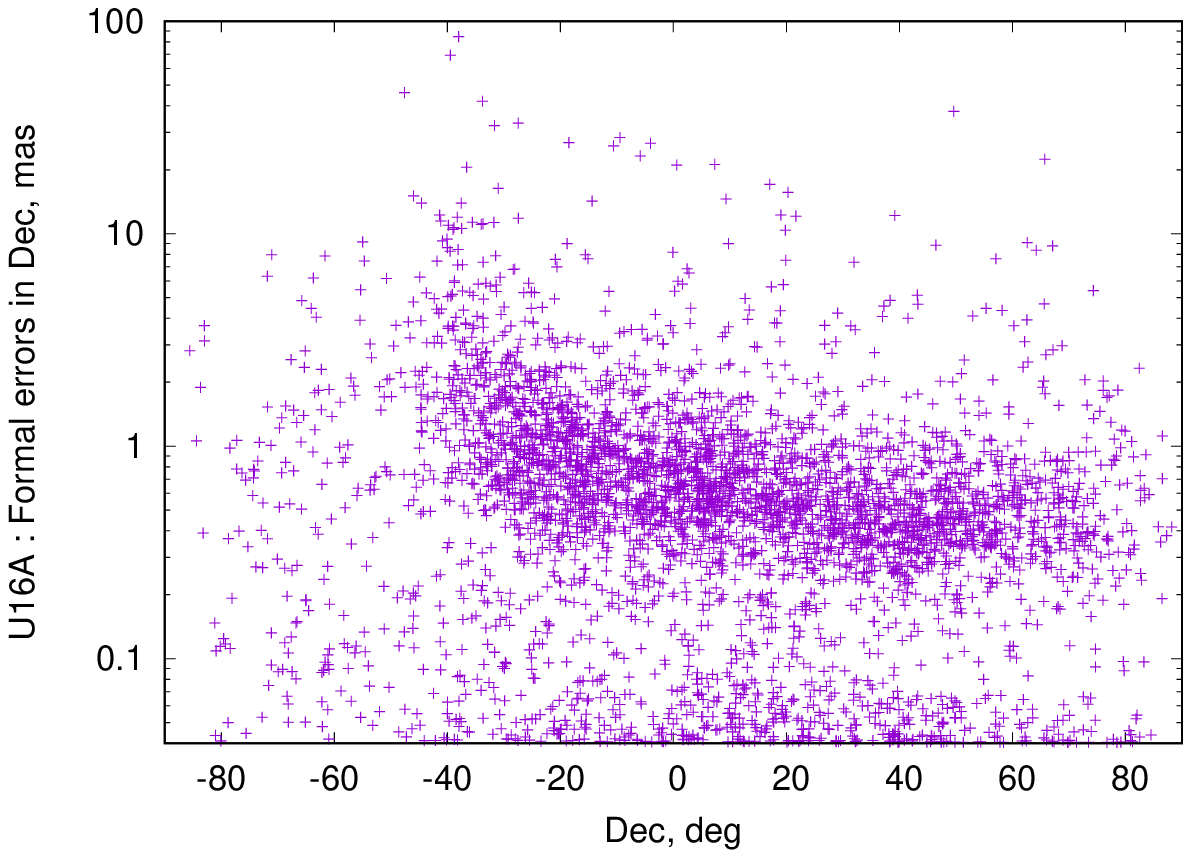}
\caption{Formal errors of U16A, inflated following formula (\ref{eq1}).}
\label{fig8}
\end{figure}

\begin{figure}[h]
\centering
\includegraphics[width=3.4in]{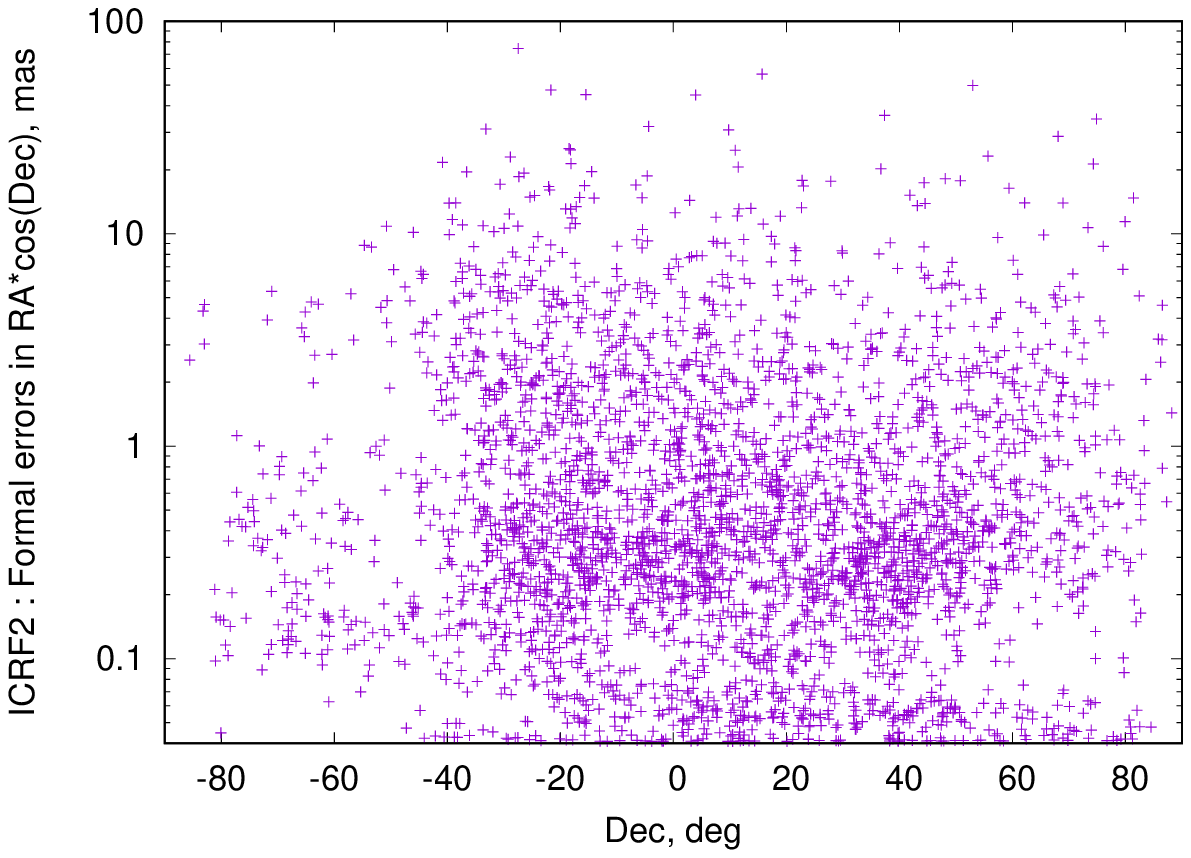}
\includegraphics[width=3.4in]{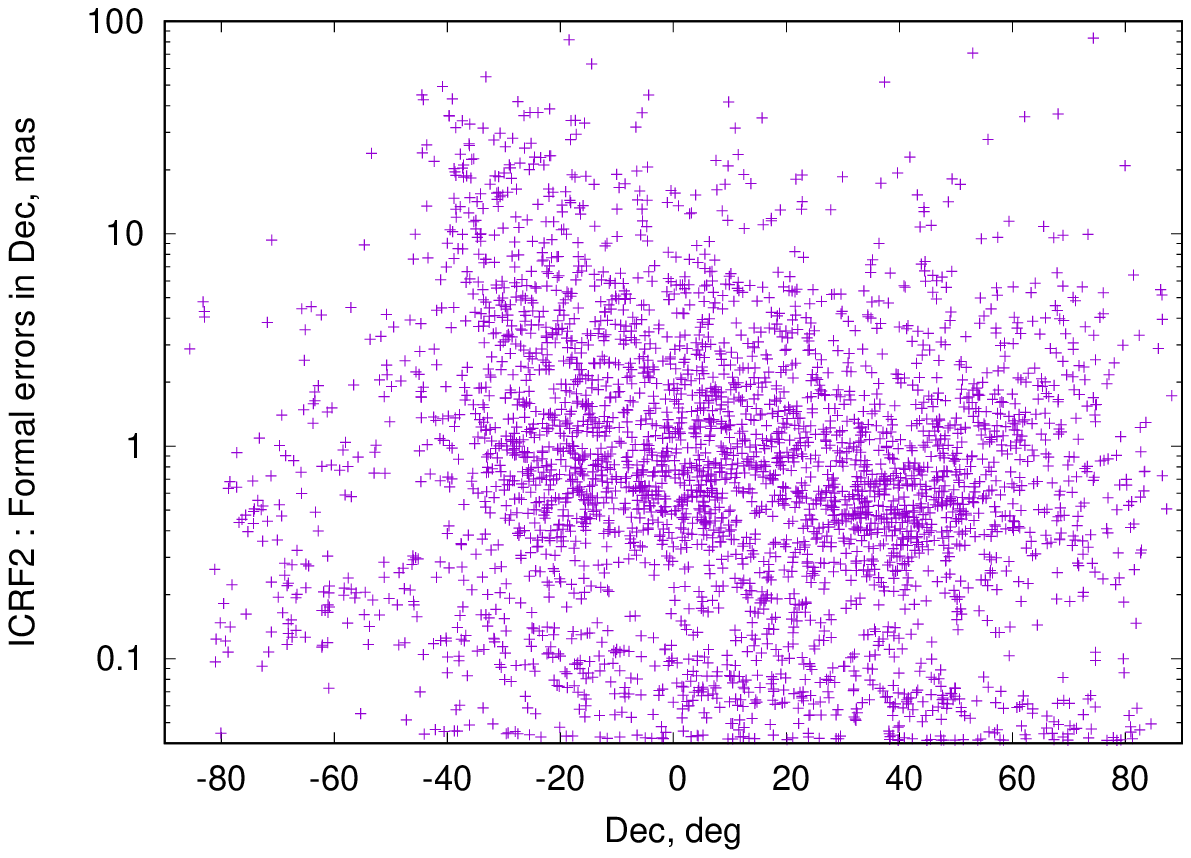}
\caption{Formal errors of ICRF2.}
\label{fig9}
\end{figure}

Since the U16A positions are improved compared to the ICRF2, the U16A formal errors should reflect this improvement and be smaller than their ICRF2 counterparts. We note that the ICRF2 formal errors were inflated (see \cite{fey+2015}), according to the formula
\begin{equation}
\sigma^2_{inflated} = (1.5 \, \sigma)^2 + (40 \, \mu as)^2 .
\label{eq1}
\end{equation}
We show the distribution of the formal errors of U16A (inflated) and ICRF2 in figures \ref{fig8} and \ref{fig9}. Approximately half of the U16A objects with inflated formal errors $<$ 0.1 mas are defining sources (and almost all the sources in the defining subset makes that cut), the other half corresponds to the 30\% most precise non-VCS sources.
Table \ref{tab9} indicates some statistics of location and scale of the formal errors. The U16A median formal errors are smaller by $\sim$ 20-25\% compared to ICRF2, while the scatter is reduced by $\sim$ 40-45\%.
\begin{table}[h]
\begin{center}
\begin{tabular}{l|l|l}
				&  Median  	&  MAD 	\\
\hline
U16A $\sigma_{\alpha}^*$ 	& 323 (312, 332)& 266 (256, 277)\\
U16A $\sigma_{\delta}$   	& 548 (530, 561)& 459 (439, 480) \\
\hline
ICRF2 $\sigma_{\alpha}^*$ 	&  397 (377, 416)	& 438 (413, 466)	\\
ICRF2 $\sigma_{\delta}$ 	&  739 (705, 773)	& 832 (788, 874) 	\\
\end{tabular}
\end{center}
\caption{Statistics of the formal errors of U16A (inflated) and ICRF2 in $\mu$as, with bootstrap confidence intervals with 95\% coverage. The MAD represents the scatter of the formal errors around their median.}
\label{tab9}
\end{table} 
Note that the formal errors in declination are significantly larger than their counterparts in right ascension (both in ICRF2 and U16A). Because of the larger number of VLBI antennas in the Northern Hemisphere, there are many more (and longer) east-west VLBI baselines than there are north-south VLBI baselines. The larger number of observations on east-west baselines is thus reflected by the smaller uncertainties in right ascension.

\begin{figure}[h]
\centering
\includegraphics[width=3.2in]{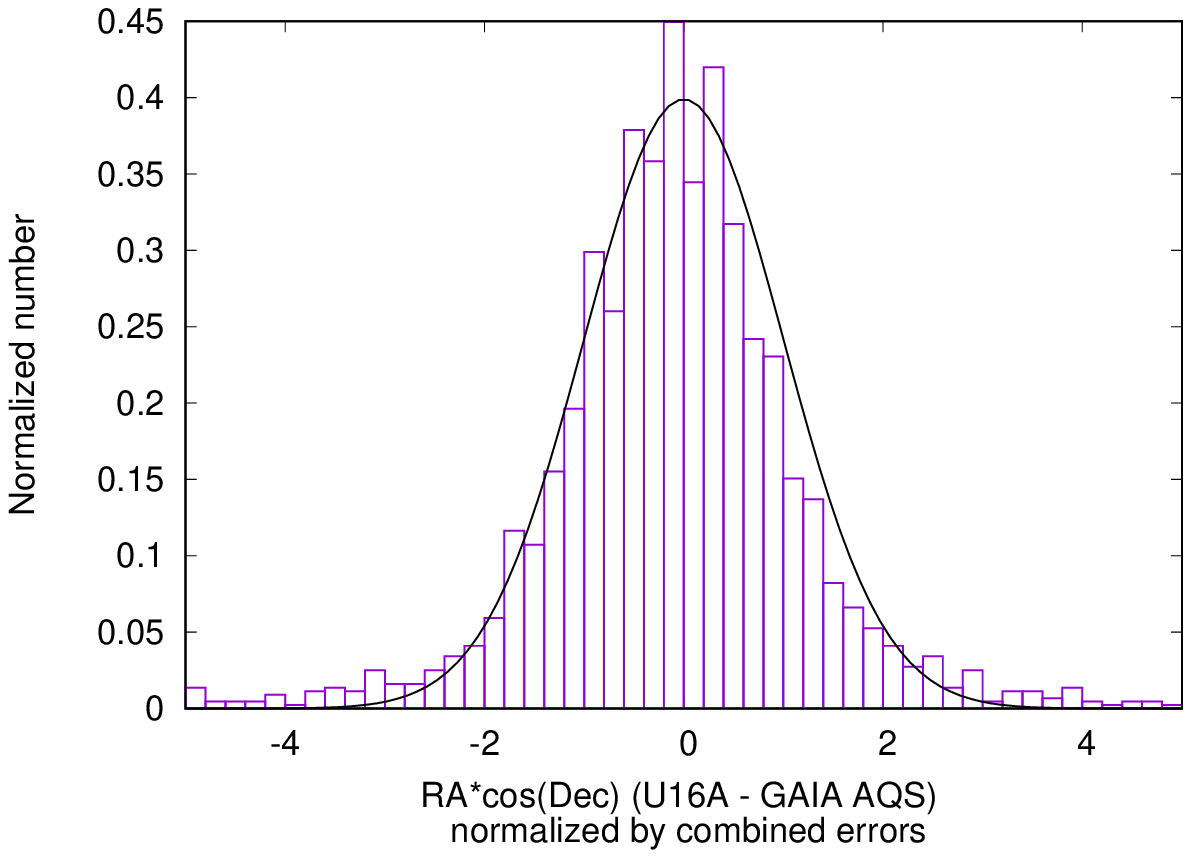}
\includegraphics[width=3.2in]{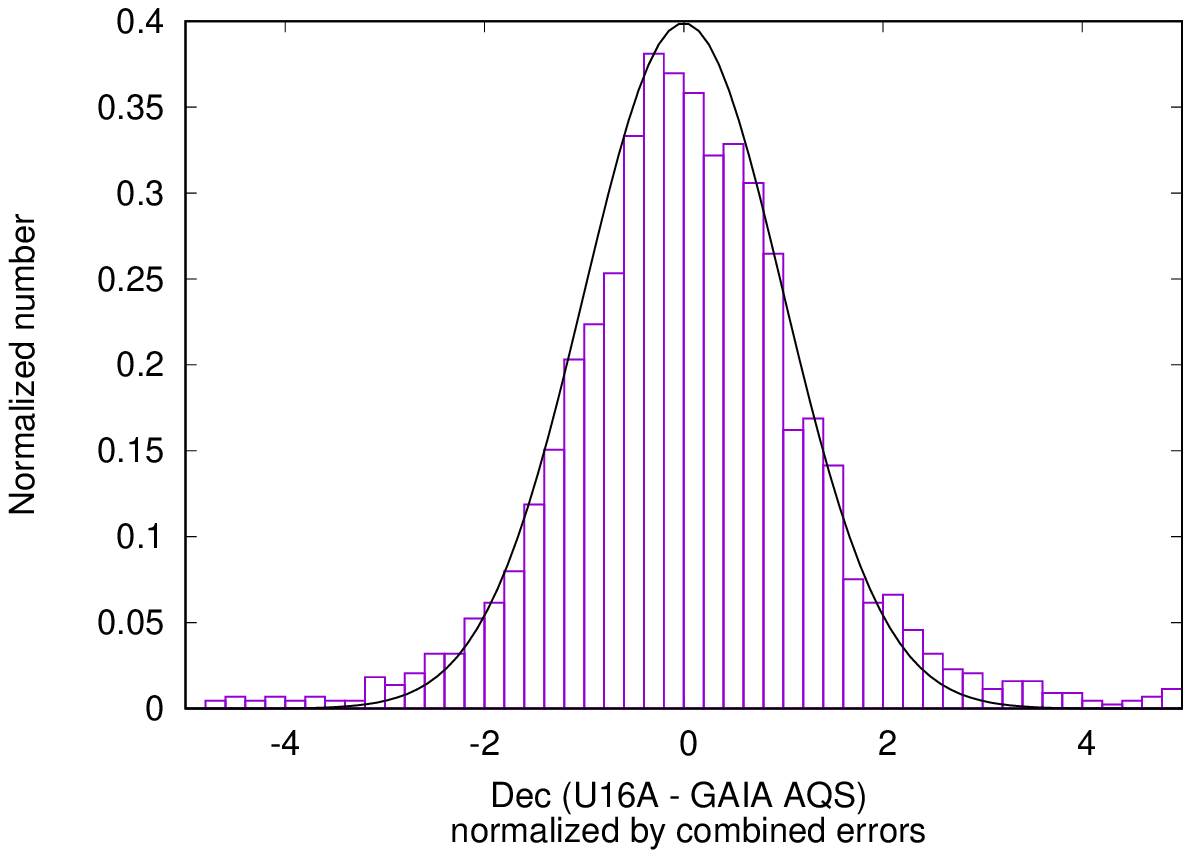}
\caption{Normalized difference U16A - Gaia AQS. A normal distribution $\mathcal{N}(0,1)$ is shown for comparison.}
\label{fig4}
\end{figure}

We show the differences (U16A - Gaia AQS) normalized by their combined errors, i.e.,
\begin{equation}
\frac{     (\alpha_{U} - \alpha_{G}) \cos \delta    }{     (\sigma^2_{\alpha^*_U} + \sigma^2_{\alpha^*_G})^{1/2}     } 
= \frac{     \alpha_{U} - \alpha_{G}   }{     (\sigma^2_{\alpha_U} + \sigma^2_{\alpha_G})^{1/2}     } , 
\, \, \, \,\,\,\,\,\,\,
\frac{     \delta_{U} - \delta_{G}   }{     (\sigma^2_{\delta_U} + \sigma^2_{\delta_G})^{1/2}     } 
\end{equation}
for both coordinates in figure \ref{fig4}. 
Both distributions are similar to normal distributions, as expected. Estimations of the scale of the distributions are $\sigma$ = 1.071 (computed when limited to the interval [-3.5,3.5]), MAD = 1.026 for RA, and $\sigma$ = 1.114, MAD = 1.105 for Dec. The formal errors of U16A in figure \ref{fig4} are not inflated. If they are inflated, we now have $\sigma$ = 1.022, MAD = 0.945 for RA, and $\sigma$ = 1.021, MAD = 0.978 for Dec. Though there are also a small number of large differences not accounted for by the formal errors, we conclude that they --- in both U16A and Gaia AQS --- broadly reflect the distribution of real errors for the majority of objects. 

\subsection{Global rotation and glide between the different frames}

We applied the VSPH expansion to investigate the global rotation and glide between the U16A, Gaia AQS and ICRF2 frames. The results are shown in table \ref{tab10}. 

Concerning the ICRF2-Gaia AQS offsets, our results are globally in agreement with the values found by \cite{mignard+2016}. We attribute the differences in the rotation and glide components listed in their table 2 and shown in their figure 14 --- which are within $\sim$ 20 $\mu$as from ours --- to the different $l_{max}$, and to the sample selection, caused by different approaches for discarding outliers. In particular we recover the significant glide in the y and z-components between the ICRF2 and Gaia (which holds for all the $l_{max}$ considered). The significant 50 $\mu$as global rotation is a weaker result, supported only by $l_{max}=1,2$.

The statistically significant components of the U16A - Gaia AQS offsets are also weaker, in that they are not supported by all the $l_{max}$ considered in this study. If we accept that the 60-70 mas glides in the y and z-components are significant, they are notably smaller than their counterparts in the ICRF2 - Gaia AQS comparison, which corroborate the results found earlier in this paper.

\section{Comparison of the new sources in U16A with Gaia DR1}

\begin{figure}[h]
\centering
\includegraphics[width=4in]{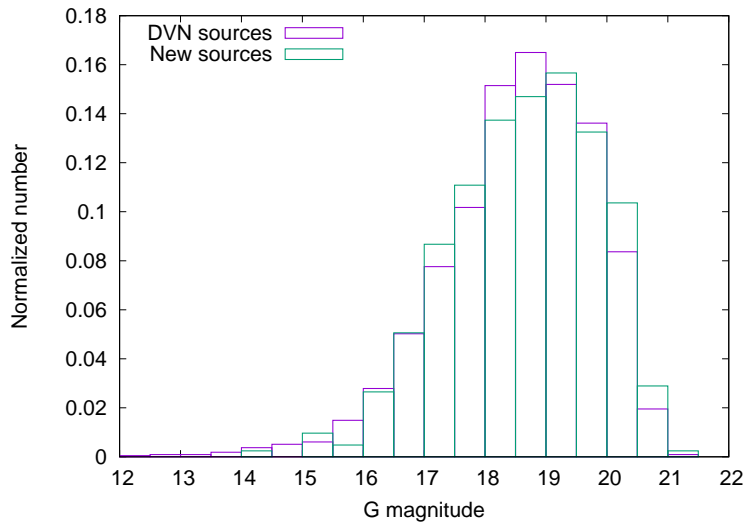}
\caption{Distribution of the Gaia G magnitude of the DVN sources (defining, VCS and non-VCS) and the new U16A sources matched with Gaia DR1.}
\label{fig1}
\end{figure}

In this section we turn our attention to the subset of 718 new sources included in U16A. These have not been previously compared to the Gaia positions. See table \ref{tab5} for more details on these sources. We crossmatched the positions of the new U16A sources with the Gaia DR1 main catalog with a 1 arcsec radius, and obtained 425 unique matches (59\% of the new U16A sources subset). The G magnitude distribution of these new sources is shown in figure \ref{fig1} and is similar to, although somewhat fainter than, the rest of the sources.

\begin{figure}[h]
\centering
\includegraphics[width=4in]{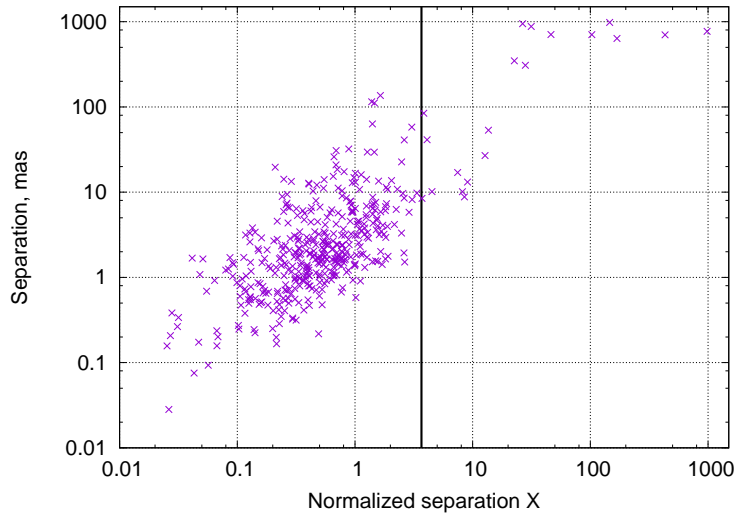}
\caption{Separation U16A-Gaia DR1 versus the normalized separation $X$ for the new sources. The U16A errors have been inflated (see text). The vertical solid line demarcates the 3.67 statistically significant normalized separation X, thus, all sources to the right of this line are considered outliers.}
\label{fig6}
\end{figure}

In order to compare the U16A - Gaia DR1 position differences with their formal errors, we computed the separation and the normalized separation $X$, as in \cite{mignard+2016}, taking into account the correlation between right ascension and declination in both U16A and Gaia DR1. They are shown in figure \ref{fig6}. Since the errors are normally distributed, $X$ should have a Rayleigh distribution. In that case, the statistically expected number of sources with $X > 3.67$ in a sample of size 425 is less than 0.5. We find 43 sources (10\% of the sample) larger than this value. If we inflate the U16A errors, we now obtain 20 outliers (5\% of the sample). This is similar to the results of \cite{mignard+2016} with regards to the position differences between ICRF2 and Gaia AQS. Half of the outliers have offsets larger than 150 mas.

\subsection{PanSTARRS images}

The sources presenting large position differences between ICRF2 and Gaia DR1 were recently investigated in detail by \cite{makarov2017} using PanSTARRS images. Those large radio-optical differences can frequently be explained by the presence of close companions and extended structures, although a number of sources defy those explanations.

We looked at the PanSTARRS images of all the new 20 sources with normalized separation $>$ 3.67. %This set includes the 22 sources with offsets $<$ 150 mas and the 10 sources with offsets $>$ 150 mas. 
The results are shown in table \ref{tab8} in Appendix B, while figure \ref{fig10} displays three sources representative of the classification in table \ref{tab8}: pair, extended, and no discernible feature. Of the 18 sources present in the PanSTARRS 3pi survey, about two thirds of them show distinctive features helping to explain the large offsets. These features are mostly the presence of a companion to the source (either true companion or chance alignement with a field star) and extended objects. Both features displace the optical photocenter detected by Gaia.

\begin{figure}[h]
\centering
\includegraphics[width=2in]{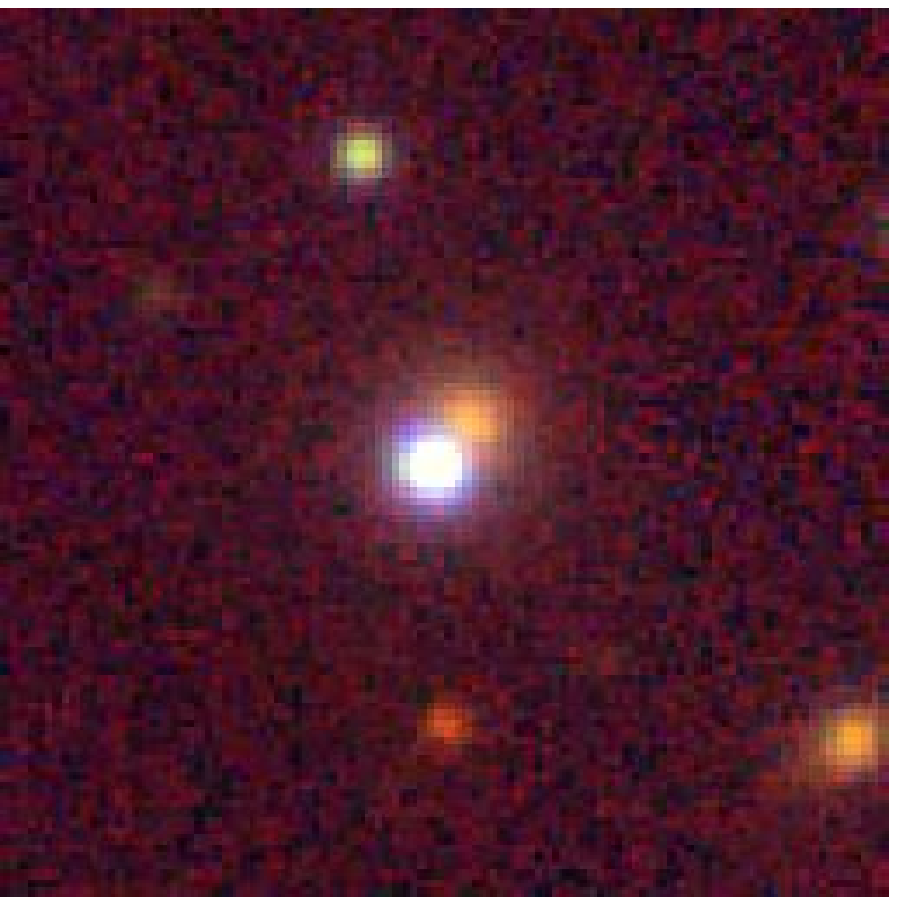}
\includegraphics[width=2in]{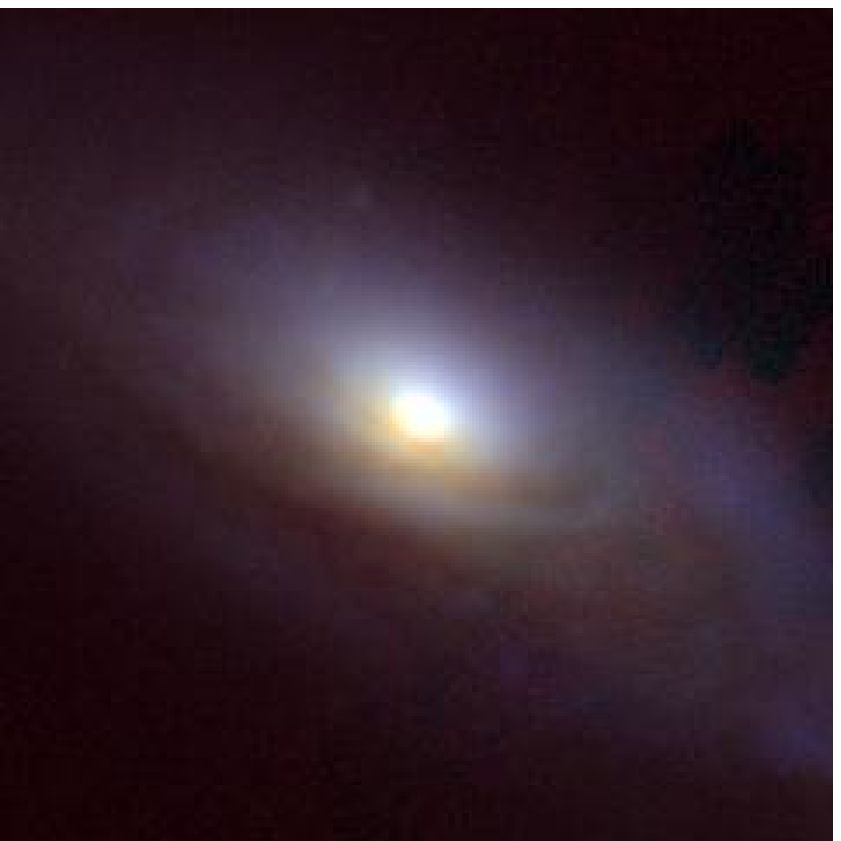}
\includegraphics[width=2in]{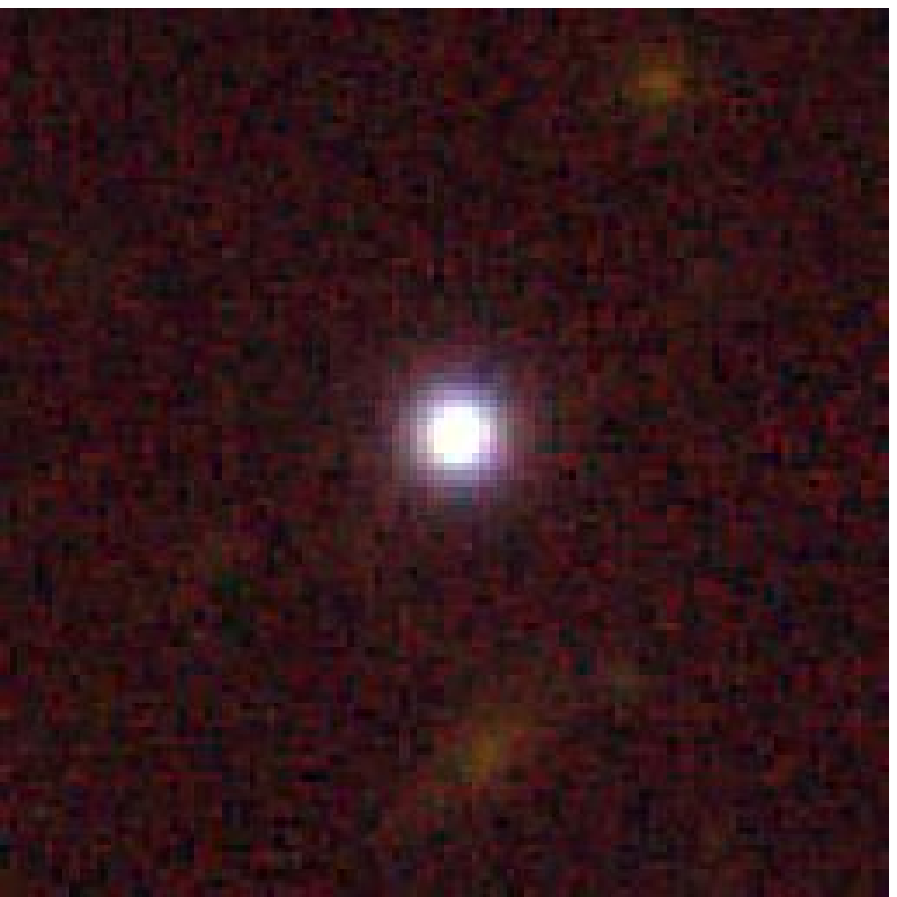}
\caption{PanSTARRS images of some of the objects in table \ref{tab8}. Left: IVS 1750+093, offset = 948 mas. Middle: NGC 5635, offset = 27 mas. Right: IVS 1117-248, offset = 309 mas. The images are 25 arcsec (left and right) and 60 arcsec wide (middle) and are centered on the U16A radio positions.}
\label{fig10}
\end{figure}

\section{Conclusions}

We compared the current USNO ICRF solution (U16A) to the ICRF2 and Gaia catalogs. The U16A solution shows some significant differences of size 0.1 mas, and smaller, relative to ICRF2, although no significant global rotation is found between these two frames. While the first Gaia release does not have the precision necessary to resolve these systematic differences in detail, the U16A solution is found to be significantly closer to Gaia than ICRF2 in both coordinates. As mentioned above, those systematic differences in declination observed in the Southern Hemisphere might be caused by observations from some of the VLBI AUST stations since 2010. However, these affect mainly the positions of Southern sources. Since the improvement of the U16A is mostly due to the greater accuracy of its VCS subset --- which is the largest component of the catalog, and is disproportionally distributed in the Northern Hemisphere and thus less influenced by the systematic differences --- it is not surprising that U16A is still found to be globally more accurate than ICRF2. Finally, with the help of PanSTARRS images, we investigated the large radio-optical offsets that some of the new U16A sources display and found an explanation for two-thirds of them.

\medskip
\medskip
\medskip
\medskip
We thank the reviewer for providing comments that improved the quality of the paper. This work also benefited from discussions and comments from Christopher Dieck, Christopher S. Jacobs, and Norbert Zacharias. This work has made use of data from the European Space Agency (ESA) mission Gaia (http://www.cosmos.esa.int/gaia), processed by the Gaia Data Processing and Analysis Consortium (DPAC, http://www.cosmos.esa.int/web/gaia/dpac/consortium). Funding for the DPAC has been provided by national institutions, in particular the institutions participating in the Gaia Multilateral Agreement. The Pan-STARRS1 Surveys (PS1) have been made possible through contributions of the Institute for Astronomy, the University of Hawaii, the Pan-STARRS Project Office, the Max-Planck Society and its participating institutes, the Max Planck Institute for Astronomy, Heidelberg and the Max Planck Institute for Extraterrestrial Physics, Garching, The Johns Hopkins University, Durham University, the University of Edinburgh, Queen’s University Belfast, the Harvard-Smithsonian Center for Astrophysics, the Las Cumbres Observatory Global Telescope Network Incorporated, the National Central University of Taiwan, the Space Telescope Science Institute, the National Aeronautics and Space Administration under Grant No. NNX08AR22G issued through the Planetary Science Division of the NASA Science Mission Directorate, the National Science Foundation under grant No. AST-1238877, the University of Maryland, and E\"{o}tv\"{o}s Lor\'{a}nd University (ELTE), and the Los Alamos National Laboratory. We made use of the Department of Defense Celestial Database of the USNO Astrometry Department. This research has made use of the VizieR catalog access tool, CDS, Strasbourg, France. The original description of the VizieR service was published in A\&AS, 143, 23.

\appendix

\section{Robust non-parametric regression}

The mean curves in figure \ref{fig3} were obtained by computing the robust equivalent of the Nadaraya-Watson estimator \citep{hardle1988,hall1990}. The classic (non-robust) Nadaraya-Watson estimator \citep{nadaraya1964,watson1964,takezawa2005} is a local weighted constant defined by 
\begin{equation}
\hat{m}(x) = 
\frac{\sum_{i=1}^N  K(\frac{x-x_i}{h}) Y_i    }{\sum_{i=1}^N  K(\frac{x-x_i}{h})} \, ,
\end{equation}
where $x$ and $Y$ are the explanatory and dependent variables, respectively, and $K(u)$ is an appropriate smoothing kernel (here we used a Gaussian distribution). This estimator is the solution of the least squares problem 
\begin{equation}
\mini_{m(x)} = \sum_{i=1}  K \bigg(\frac{x-x_i}{h} \bigg) (Y_i - m(x))^2 \text{ \hspace{0.5cm} or, equivalently, \hspace{0.5cm}} \sum_{i=1}  K \bigg(\frac{x-x_i}{h} \bigg) (Y_i - m(x)) = 0 \, .
\end{equation}
The bandwidth $h$ is a free parameter and determines whether the procedure overfits or underfits the real offsets.

A robust M-estimator \citep{huber2009} for $m(x)$ is obtained by replacing the square with a robust function $\psi(u)$ and solving
\begin{equation}
\sum_{i=1}  K \bigg(\frac{x-x_i}{h} \bigg) \psi \bigg(\frac{Y_i - m(x)}{s} \bigg) = 0 \, ,
\end{equation}
where $s$ is a robust scale estimate. We used the Huber function
\begin{equation}
\begin{split}
 \psi(u) & = -1   		\text{\hspace{0.1cm},\hspace{1cm}} u \leq -c \\
 \psi(u) & = \frac{u}{c}    	\text{\hspace{0.1cm},\hspace{1cm}} \,\, -c < u \leq c  \\
 \psi(u) & = 1    		\text{\hspace{0.1cm},\hspace{1cm}} \,\,\,\, u > c 
\end{split}
\end{equation}
where the coefficient $c=1.345$ is chosen so as to give a 95\% efficiency in case the errors follow a simple Gaussian distribution \citep{holland1977}. 

The estimate $\hat{m}(x)$ is obtained as follows:
\begin{enumerate}
 \item Compute a local scale $s$, by computing the MAD of the $Y_i$ values in an interval around $x$ (say 10 degrees)
 \item Solve iteratively for $m(x)$ with Newton's method until a satisfactory convergence is reached
 \begin{equation}
  m(x)^{(p+1)} = m(x)^{(p)} + \frac{\sum_{i=1}^N  K(\frac{x-x_i}{h})  \psi (\frac{Y_i - m(x)^{(p)}}{s} )      }{\sum_{i=1}^N  K(\frac{x-x_i}{h})  \psi^{\prime} (\frac{Y_i - m(x)^{(p)}}{s} )} \, s \, ,
 \end{equation}
where the initial guess  $m(x)^{(0)}$ is the local median of the $Y_i$ values.
\end{enumerate}

The choice of the bandwidth $h$ is crucial and is usually computed by using cross-validation, i.e. by finding $h$ that minimizes
\begin{equation}
CV(h) =  \sum_{i=1}^N |Y_i - \hat{m}(x)^{(-i)} | \, ,
\end{equation}
where $\hat{m}(x)^{(-i)}$ is the estimate computed by omitting the i$^{th}$ source. Here the cross-validation step is robustified by using a sum of absolute values instead of the usual sum of squares \citep{leung2005}.

The curves corresponding to the confidence intervals in figure \ref{fig3} were computed using wild bootstrap resampling \citep{liu1988,mammen1993}. The non-parametric regression curve, computed as described above, gives the residuals $r_i = Y_i - \hat{m}(x_i)$. The $Y_i$ values of the wild bootstrap samples are called $Y^*_i$ and are computed following
\begin{equation}
Y^*_i = Y_i + \varepsilon_i \,  r_i \, ,
\end{equation}
where the $\varepsilon_i$ follow a Rademacher distribution (see \cite{flachaire2005})
\begin{equation}
\begin{split}
 \varepsilon_i & = -1   \text{\hspace{0.1cm},\hspace{1cm}} p = 0.5 \\
 \varepsilon_i & =  1   \text{\hspace{0.1cm},\hspace{1cm}} \,\,\,\, p = 0.5  \, .
\end{split}
\end{equation}
This amounts to randomly changing the sign of the residuals with respect to the mean curve.

A number of wild bootstrap samples (200) were computed for each mean curve, and the estimator $\hat{m}(x)$ was computed for each of the samples. The distribution of the resulting estimators $\hat{m}^*_{B}(x)$ with $B=1,...,200$ is then used to compute a 95 \% confidence interval at each point $x_i$.

\section{U16A new sources with statistically significant radio-optical offsets with Gaia DR1}

\begin{sidewaystable}
\begin{center}
\footnotesize
\begin{tabular}{llrrrlrrrrr}
IVS name   & Gaia DR1 source id		& U16A RA 	& U16A Dec 		& U16A-Gaia  	&Features in PS1 image, 				& Gaia DR1 		& \multicolumn{3}{c}{Gaia DR1 (mean / max} 		&Number of \\
	   &				&		&			& offset	&close Gaia neighbors and notes				& excess noise	 	& \multicolumn{3}{c}{of excess noise / \# of} 	&VLBI \\
	   &				&		&			& 		&							& (excess noise $\sigma$)&  \multicolumn{3}{c}{neighbors) within 0.1 deg}	&sessions\\	  
	   &				& deg 		& deg			& mas		& 							& mas			&mas&mas&			&\\		
\hline	
\hline	
\multicolumn{3}{l}{Reasonable or tentative explanation} 	 	& 			& 		& 									&			&			&\\
2048$+$370 & 1870909710012536832		& 312.711411916 & 37.2425211822 	&982.598	&Possibly double sep. 1.5”, object at 5"		&0. 	(0.)		&1.305 & 7.769 &   81	&2\\
1750$+$093 & 4488716878495401600		&  268.259853834& 9.3333374835 		&947.725	&Double sep. 1.5"					&0.757 	(3.88)		& 1.572 & 4.985 &    25	&1\\
1740$-$169 & 4123749111948788480		&  265.775909615& -16.9711020499	&772.631	&Extended, Gaia neighbor at 4"				& 0.563 (1.93)		& 1.351 & 18.129 &   329&2\\ 
1858$-$212 & 4081909331557329664		& 285.268558136 & -21.2003238993	&708.536	&Double sep. 0.5" or lens, Gaia neighbor at 4.9"	&2.276	(1.92)		&1.233& 5.650 &   116	&2\\ 
1701$-$246 & 4112708198449948160		& 256.25534233 	& -24.7527101571	&706.490	&2MASSX galaxy, Gaia neighbor at 4.7"			& 2.072 (13.53)		&1.450 & 11.605 &   175	&2\\
0931$+$103 & 589443338929476864		&  143.44211412	& 10.1524506152 	&84.398		&NGC 2911, well-resolved galaxy with dust		&15.667	(2370.78)	&1.149 & 1.580 &     2	&2\\
1151$+$126 & 3919707688292070784		&178.543363944  & 12.4193758601 	&53.367		&Galaxy							&15.451 (278.46)	&1.897 & 1.897 &     1	&1\\ 
0310$+$410 & 239542077129105152		&48.4901913843  & 41.2566596249		&41.217		&2MASSX galaxy						&10.932 (1120.03)	& 1.175 & 3.194 &    17	&1\\
NGC 5635   & 1280540719731180544		& 217.132340338 & 27.408953535		&26.889		&Well-resolved galaxy with dust				&9.511	(2032.17)	&1.992 & 1.992&     1	&2\\	
2338$-$295 & 2327748343150961408		&  355.374009923& -29.320843913 	&10.171		&Galaxy, Gaia neighbor at 3.8"				&6.893  (381.44)	&7.577 & 18.464 &    6	&3\\ 		
1638$+$118 & 4447290166579343616		& 250.245385766 & 11.734503378 		&10.124		&Galaxy							&3.147 	(125.71)	&1.270 & 2.414 &     7	&2\\		
1400$+$162 & 1231616197506959104		&  210.685473087& 15.9990737579		&8.458		&Galaxy							&1.260 (4.32)		&2.280 & 2.779 &     3	&2\\
\hline	
\multicolumn{2}{l}{No explanation} 	&  		& 			& 		& 							& 				&			&\\
1834$-$155 & 4102987622313209472		& 279.406967906 & -15.5455694377 	&880.142	&Gaia neighbors at 1.7" and 3.7"			& 0.695 (8.67)		&1.445 & 8.119 &   264	&2\\ 
1911$+$013 & 4264732100629035264		& 288.560639404 & 1.40730671431		&701.242	&							&1.784 	(20.78)		&1.775 & 18.095 &  238	&2\\
1143$+$052 & 3896995007877942016		& 176.632255657 & 4.97202463292		&636.309	&							& 2.254 (1.09)		&1.457 & 1.614  &   2	&2\\
1117$-$248 & 3534310892614244992		&  170.038022714& -25.1354989919 	&308.826	&							&0.626 	(8.60)		&0.938 & 1.719 &     7	&5\\
1221$+$484 & 1545500550756588032		& 185.992627829 & 48.2160179775		&13.151		&							&4.171	(15.66)		& 1.006 & 1.258 &  3	&1\\
1216$+$179 & 3946753131714484992		& 184.69418527 	& 17.638130057 		&8.800		&							& 0.479 (2.52)		& & & 0			&1\\  
\hline
\multicolumn{2}{l}{Sources not in PS1 images} 	&  	& 			& 		& 							& 					&			&\\
1738$-$336 & 4053875702081678080		& 265.515372964 & -33.6928928969	&347.617	&Gaia neighbor at 4.7"					& 1.159 (2.21)			&1.361 & 5.386 &  369	&2\\
0912$-$330 & 5630916907781932928		&138.653004231	&-33.2478928246		&17.001		&							& 0. (0.)			& 0.838 & 4.475 &    23	&1\\
\end{tabular}
\end{center}
\caption{Features observed on the PanSTARRS images for the 20 sources with large separations U16A - Gaia DR1.}
\label{tab8}
\end{sidewaystable}

%%%%%%%%%%%%%%%%%%%%%%%%%%%%%%%%%%%%%%%%%%%%%%%%%%%%%%%%%%%%%%%%

\end{document}